\documentclass[10pt,journal,twocolumn]{IEEEtran}
\IEEEoverridecommandlockouts
\usepackage{cite}
\usepackage{amsmath,amssymb,amsfonts,booktabs}
\usepackage{graphicx}
\usepackage{textcomp}
\usepackage{xcolor}
\usepackage{enumerate}
\usepackage{epstopdf}
\usepackage{subcaption}
\usepackage{amsthm}

\usepackage{enumitem}
\usepackage{subfig}
\usepackage{subcaption}
\usepackage{algorithm}
\usepackage{algorithmic}
\usepackage{verbatim}

\allowdisplaybreaks[4]

\usepackage{bbm}
\usepackage{caption}

\usepackage{algorithm}

\usepackage{verbatim}
\usepackage{booktabs}
\usepackage{float}
\usepackage{makecell}
\usepackage{lettrine}

\usepackage{float}  
\usepackage{enumerate}
\usepackage{epstopdf}
\usepackage{subcaption}

\usepackage{amsfonts}

\usepackage{multirow} 
\usepackage{bm}

\usepackage{mathtools}

\usepackage{setspace}

\def\BibTeX{{\rm B\kern-.05em{\sc i\kern-.025em b}\kern-.08em
		T\kern-.1667em\lower.7ex\hbox{E}\kern-.125emX}}

\begin{document}

\title{\LARGE Latency-Aware Resource Allocation for Mobile Edge Generation and Computing via Deep Reinforcement Learning}
\author{
        Yinyu Wu,~\IEEEmembership{Student Member,~IEEE},
        Xuhui Zhang,~\IEEEmembership{Student Member,~IEEE},
        Jinke Ren,~\IEEEmembership{Member,~IEEE},\\
        Huijun Xing,~\IEEEmembership{Student Member,~IEEE},
        Yanyan Shen,~\IEEEmembership{Member,~IEEE},
        and Shuguang Cui,~\IEEEmembership{Fellow,~IEEE}



\thanks{
Y. Wu is with Shenzhen Institute of Advanced Technology (SIAT), Chinese Academy of Sciences, Guangdong 518055, China, and also with the University of Chinese Academy of Sciences, Beijing 100049, China (e-mail: yg.wu@siat.ac.cn).
}

\thanks{
X. Zhang, J. Ren, and H. Xing are with the Shenzhen Future Network of Intelligence Institute (FNii-Shenzhen), the School of Science and Engineering (SSE), and the Guangdong Provincial Key Laboratory of Future Networks of Intelligence, The Chinese University of Hong Kong, Shenzhen, Guangdong 518172, China (e-mail: xu.hui.zhang@foxmail.com; jinkeren@cuhk.edu.cn; huijunxing@link.cuhk.edu.cn).
}


\thanks{
Y. Shen is with the SIAT, Chinese Academy of Sciences, Guangdong 518055, China, and also with the Shenzhen University of Advanced Technology, Guangdong 518055, China (e-mail: yy.shen@siat.ac.cn).
}

\thanks{
S. Cui is with the SSE, the FNii-Shenzhen, and the Guangdong Provincial Key Laboratory of Future Networks of Intelligence, The Chinese University of Hong Kong, Shenzhen, Guangdong 518172, China (e-mail: shuguangcui@cuhk.edu.cn).
}

\vspace{-2.7em}
}

\maketitle

\begin{abstract}
Recently, the integration of mobile edge computing (MEC) and generative artificial intelligence (GAI) technology has given rise to a new area called
mobile edge generation and computing (MEGC), which offers mobile users heterogeneous services such as task computing and content generation.
In this letter, we
investigate the joint communication, computation, and the AIGC resource allocation problem in an MEGC system. A latency minimization problem is first formulated to enhance the quality of service for mobile users. 
Due to the strong coupling of the optimization variables, we propose a new deep reinforcement learning-based algorithm to solve it efficiently, 
Numerical results demonstrate that the proposed algorithm can achieve lower latency than several baseline algorithms.
\end{abstract}

\begin{IEEEkeywords}
Artificial intelligence, mobile edge generation and computing, deep reinforcement learning.
\end{IEEEkeywords}

\section{Introduction}
\IEEEPARstart{I}{n} recent years, artificial intelligence generative content (AIGC) has gained widespread attention due to its powerful creative ability for a variety of content, such as images, videos, and music \cite{10398474}.
Several examples, including the generative pre-trained transformer (GPT) developed by OpenAI and the WaveNet developed by DeepMind, have shown great potential in enhancing communication performance for the next-generation wireless networks \cite{10578004}.
By deploying AIGC services at the network edge,
lower latency and reduced communication overhead can be achieved for mobile users (MUs) \cite{xu2024}.
Meanwhile, computing services remain crucial due to the increasing demand for handling computationally intensive tasks for MUs \cite{8016573}.
By processing task data at the network edge, the shortcomings of network congestion and long latency in conventional mobile computing systems can be addressed
\cite{8664595, 8387798}.
Hence, the integration of communication, computing, and generation services at the network edge is promising to
address the heterogeneous requests in future wireless networks.

Several pioneering works have exploited the applications of AIGC services in mobile edge computing (MEC) systems \cite{10398474, 10472660, xu2024, xu20242}.
An edge intelligence infrastructure was proposed to provide personalized and low latency AIGC services \cite{10398474}.
To improve the user utility, a pricing-based mechanism was proposed in \cite{10472660}, which investigated the efficient AIGC services.
Moreover, a mobile edge generation (MEG) system was proposed to reduce the distributed computation and transmission overhead \cite{xu2024}.
Furthermore, an MEG enabled digital twin system was studied in \cite{xu20242}, which can be applied in both single-user and multi-user scenarios.
Besides, the heterogeneous services in MEC systems were investigated in previous works \cite{10319405, 10535750, 10606316}.
A three-stage heterogeneous computing model was proposed in \cite{10319405} to practically describe the computation process of parallelizable tasks.
To maximize the computation efficiency of a multi-server system, an advantage-actor-critic-based deep reinforcement learning (DRL) method was proposed in \cite{10535750}.
In \cite{10606316}, an unmanned aerial vehicle assisted heterogeneous MEC system was studied, where the MUs chose different service providers to maximize the task computation volume.
Additionally, to minimize the average latency and improve the MU's quality of experience,
the joint optimization of data offloading, resource allocation, and data caching time are investigated in \cite{9557844}.

Most previous works focus on the computing resource assignments or the AIGC service allocations in MEC and MEG, while the integration of MEC and MEG to provide heterogeneous services for MUs has not been exploited.
To provide heterogeneous computation, AIGC, and vision enhancement (VE) services for MUs while improving the user experience, we aim to minimize the average latency of all MUs in a novel mobile edge generation computing (MEGC) system. 
A DRL-based latency-aware resource allocation algorithm is proposed to jointly optimize the bandwidth allocation, the backhaul transmit power, the computation resources, and the task offloading ratio.
The superior latency performance of the proposed algorithm is verified by comparing to benchmark algorithms.

\section{System Model and Problem Formulation}
\begin{figure}[h!]
	\centering
	\includegraphics[width=1.705in]{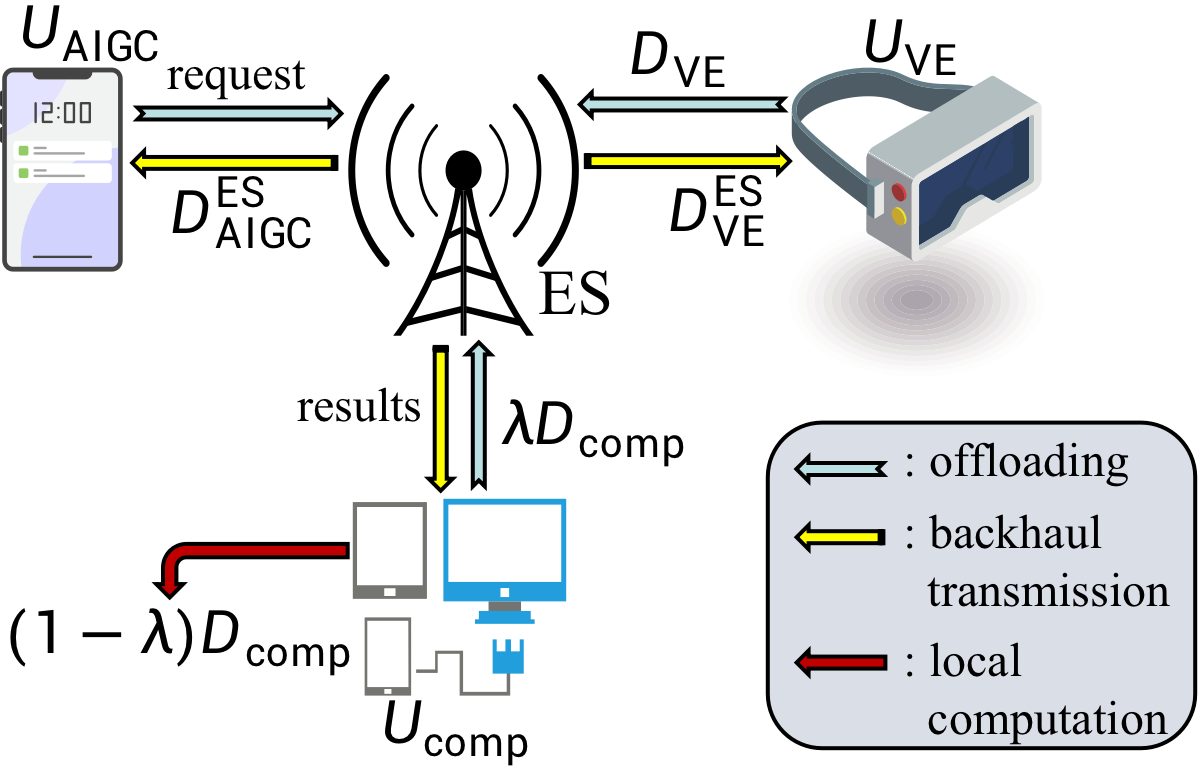}
	\caption{MEGC system model.}
	\label{fig:system}
\end{figure}

To enable the heterogeneous task requirements, we consider an MEGC system that can provide both computation services and AIGC services, as illustrated in Fig. \ref{fig:system}.
The MEGC system consists of an edge server (ES) equipped with a powerful computing and generating unit, and three kinds of MUs with different task requests, including computation, AIGC, and VE, respectively\footnote{
{The MEGC system can be directly extended to the system with multiple MUs, where the MUs with identical tasks can be aggregated as a cluster.
}}.
et $U_{\mathsf{comp}}$, $U_{\mathsf{AIGC}}$, $U_{\mathsf{VE}}$ denote the MUs with computation task, AIGC task, and VE task, respectively. The serving time is index by $\mathcal{T} = \{1,\cdots, T\}$.
Each MU requests one service during each time slot, and the frequency division multiple access (FDMA) technology is adopted in each time slot of request turn.The system model is shown in Fig \ref{fig:system}.

\subsection{Communication Model}
\begin{figure*}[h!]
	\centering
	\includegraphics[width=5.25in]{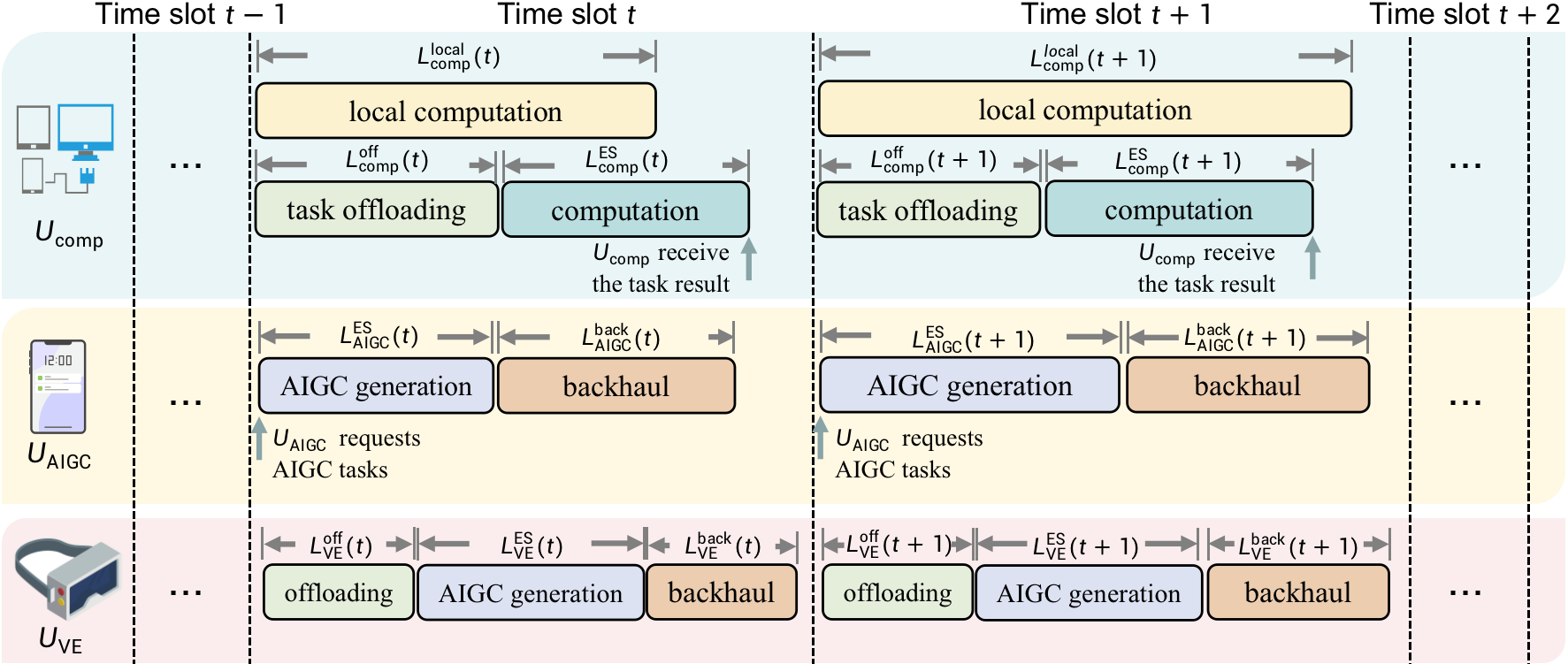}
	\caption{Two stages of data offloading and result backhaul transmission in MEGC.}
	\label{Decision process}
\end{figure*}

As illustrated in Fig. \ref{Decision process}, the MEGC system consists of two stages: data offloading and result backhaul transmission. In the data offloading stage, the MUs $U_{\mathsf{comp}}$ and $U_{\mathsf{VE}}$ need to offload their data to the ES, while the MU $U_{\mathsf{AIGC}}$ only needs to offload a request to the ES.
During the backhaul transmission stage, the ES transmits the generated content and processed images/videos to the MUs $U_{\mathsf{AIGC}}$ and $U_{\mathsf{VE}}$, respectively,
and transmits the computing results to the MU $U_{\mathsf{comp}}$.

Since the FDMA technology is adopted, the bandwidth resource is optimized to allocate for different MUs within each time slot. Let $B_{\mathsf{off}}$ denote the bandwidth allocated to the offloading stage. Thus, the offloading rates of the MU $U_{\mathsf{comp}}$ and $U_{\mathsf{VE}}$ at time slot $t$ are expressed respectively as
\begin{equation}
    r_{\mathsf{comp}}^{\mathsf{off}} (t) = \alpha_{\mathsf{comp}}^{\mathsf{off}}(t) B_{\mathsf{off}} \log_2
    \left ( 1 + \frac{p_{\mathsf{comp}} h_{\mathsf{comp}}^{\mathsf{off}} (t)}{\alpha_{\mathsf{comp}}^{\mathsf{off}}(t) B_{\mathsf{off}}N_0}
    \right ),
\end{equation}
\begin{equation}
    r_{\mathsf{VE}}^{\mathsf{off}} (t) = \alpha_{\mathsf{VE}}^{\mathsf{off}}(t) B_{\mathsf{off}} \log_2
    \left ( 1 + \frac{p_{\mathsf{VE}} h_{\mathsf{VE}}^{\mathsf{off}} (t)}{\alpha_{\mathsf{VE}}^{\mathsf{off}}(t) B_{\mathsf{off}}N_0}
    \right ),
\end{equation}
where $\alpha_{\mathsf{comp}}^{\mathsf{off}}(t) \in [0,1]$ and $\alpha_{\mathsf{VE}}^{\mathsf{off}}(t) \in [0,1]$ are the bandwidth allocation ratios of the MUs $U_{\mathsf{comp}}$ and $U_{\mathsf{VE}}$, which satisfy $\alpha_{\mathsf{comp}}^{\mathsf{off}}(t) + \alpha_{\mathsf{VE}}^{\mathsf{off}}(t) = 1, \forall t$. $p_{\mathsf{comp}}$ and $p_{\mathsf{VE}}$ denote the transmit power the MUs $U_{\mathsf{comp}}$ and $U_{\mathsf{VE}}$, respectively.
$h_{\mathsf{comp}}^{\mathsf{off}} (t)$ represents the line-of-sight (LoS) channel gain between the MU $U_{\mathsf{comp}}$ and the ES, and $h_{\mathsf{VE}}^{\mathsf{off}} (t)$ is the LoS channel gain between the MU $U_{\mathsf{VE}}$ and the ES.
$N_0$ is the power spectral density.

Similarly, let $B_{\mathsf{back}}$ denote the bandwidth allocated to the backhaul transmission stage. Then, the backhaul transmission rates of the MUs $U_{\mathsf{AIGC}}$ and $U_{\mathsf{VE}}$ are given by
\begin{equation}
    r_{\mathsf{AIGC}}^{\mathsf{back}} (t) = \alpha_{\mathsf{AIGC}}^{\mathsf{back}}(t) B_{\mathsf{back}} \log_2
    \left ( 1 + \frac{\beta(t)P_{\mathsf{}} h_{\mathsf{AIGC}}^{\mathsf{back}} (t)}{\alpha_{\mathsf{AIGC}}^{\mathsf{back}}(t) B_{\mathsf{back}}N_0}
    \right ),
\end{equation}
\begin{equation}
    r_{\mathsf{VE}}^{\mathsf{back}} (t) = \alpha_{\mathsf{VE}}^{\mathsf{back}}(t) B_{\mathsf{back}} \log_2
    \left ( 1 + \frac{(1-\beta(t))P_{\mathsf{}} h_{\mathsf{VE}}^{\mathsf{back}} (t)}{\alpha_{\mathsf{VE}}^{\mathsf{back}}(t) B_{\mathsf{back}}N_0}
    \right ),
\end{equation}
where $\alpha_{\mathsf{AIGC}}^{\mathsf{back}}(t) \in [0,1]$ and $\alpha_{\mathsf{VE}}^{\mathsf{back}}(t) \in [0,1]$ are the bandwidth allocation ratios of the MUs $U_{\mathsf{AIGC}}$ and $U_{\mathsf{VE}}$, which satisfy $\alpha_{\mathsf{AIGC}}^{\mathsf{back}}(t) + \alpha_{\mathsf{VE}}^{\mathsf{back}}(t) = 1, \forall t$.
$h_{\mathsf{AIGC}}^{\mathsf{back}} (t)$ represents the LoS channel gain between the ES and the MU $U_{\mathsf{AIGC}}$, and $h_{\mathsf{VE}}^{\mathsf{back}} (t)$ is the LoS channel gain between the ES and the MU $U_{\mathsf{VE}}$.
$P$ is the transmit power of the ES.
$\beta_{\mathsf{}}^{\mathsf{}}(t) \in [0,1]$ 
is the power allocation ratio to the ES for communicating with the MU $U_{\mathsf{AIGC}}$, while $1-\beta_{\mathsf{}}^{\mathsf{}}(t)$ 
is the power allocation ratio for the ES to communicate with the MU $U_{\mathsf{VE}}$.

\subsection{Latency Model}
\paragraph{Latency of MU $U_{\mathsf{comp}}$}
The MU $U_{\mathsf{comp}}$ requests data computing services. At the beginning of each time slot, a data packet is arrived with data volume $D_{\mathsf{comp}}$, which follows a Poisson distribution with density $\rho_{\chi}$ Mbits.
Assume that the data packet can be partially computed by its local processor and partially offloaded to the ES for computing. Let $\lambda (t)$ denote the ratio of the data packet offloaded to the ES at time slot $t$. Then, the offloading latency is given by
\begin{equation}
    L_{\mathsf{comp}}^{\mathsf{off}} (t) = \frac{\lambda(t) D_{\mathsf{comp}} (t)}{r_{\mathsf{comp}}^{\mathsf{off}} (t)}.
\end{equation}
Assume the local computing capability of the MU $U_{\mathsf{comp}}$ is $f_{\mathsf{comp}}$ central processing unit (CPU) cycles per second.
The local computing latency is thus represented by
\begin{equation}
    L_{\mathsf{comp}}^{\mathsf{local}} (t) = \frac{(1-\lambda(t))\chi D_{\mathsf{comp}}}{f_{\mathsf{comp}}},
\end{equation}
where $\chi$ denotes the number of CPU cycles for computing one bit data. Meanwhile, the computing latency of the ES for processing the data offloaded by the MU $U_{\mathsf{comp}}$ at time slot $t$ is given by
\begin{equation}
    L_{\mathsf{comp}}^{\mathsf{ES}} (t) = \frac{\lambda(t)\chi D_{\mathsf{comp}}}{\omega_{\mathsf{comp}} (t) f_{\mathsf{ES}}},
\end{equation}
where $f_{\mathsf{ES}}$ is the maximum computing capability of the ES, and $\omega_{\mathsf{comp}} (t)$ denotes the ratio of the computation resources allocated to the MU $U_{\mathsf{comp}}$ at time slot $t$.

The overall latency of the MU $U_{\mathsf{comp}}$ is the maximum of the latency between local computing and ES computing, which is derived as
\begin{equation}
    L_{\mathsf{comp}} (t) = \max (L_{\mathsf{comp}}^{\mathsf{local}} (t) , L_{\mathsf{comp}}^{\mathsf{off}} (t)+L_{\mathsf{comp}}^{\mathsf{ES}} (t)).
\end{equation}

\paragraph{Latency of MU $U_{\mathsf{AIGC}}$}
The MU $U_{\mathsf{AIGC}}$ requests AIGC services. Since the data volume of the request is very small, the time latency for transmitting the request can be ignored. After receiving the request, the ES begins to generate results.
The time latency for AIGC inference is given by
\begin{equation}
    L_{\mathsf{AIGC}}^{\mathsf{ES}} (t) = \frac{\xi \chi D_{\mathsf{AIGC}}^{\mathsf{ES}} (t) + \zeta}
    {\omega_{\mathsf{AIGC}}(t)f_{\mathsf{ES}}},
\end{equation}
where $D_{\mathsf{AIGC}}^{\mathsf{ES}} (t)$ is the data volume of the expected AIGC results, related to the image resolution. $\xi$ is the coefficient related to the ES-embedded LLM and computing hardware, and $\zeta$ is the coefficient related to the minimum computing resources required for a small-scale image inference at the ES.
Besides, $\omega_{\mathsf{AIGC}} (t)$ denotes the ratio of the computation resources allocated to the MU $U_{\mathsf{AIGC}}$ at time slot $t$.
Accordingly, the latency for transmitting the AIGC results to the MU $U_{\mathsf{AIGC}}$ can be expressed by
\begin{equation}
    L_{\mathsf{AIGC}}^{\mathsf{back}} (t) = \frac{D_{\mathsf{AIGC}}^{\mathsf{ES}} (t)}{r_{\mathsf{AIGC}}^{\mathsf{back}} (t)}.
\end{equation}

Hence, the overall latency of the MU $U_{\mathsf{AIGC}}$ is given by
\begin{equation}
    L_{\mathsf{AIGC}}^{\mathsf{}} (t) = L_{\mathsf{AIGC}}^{\mathsf{ES}} (t) + L_{\mathsf{AIGC}}^{\mathsf{back}} (t).
\end{equation}

\paragraph{Latency of MU $U_{\mathsf{VE}}$}
The MU $U_{\mathsf{VE}}$ requests VE services, i.e., the image and/or video processing services. One service consists of three parts: the original data offloading, the data inference at the ES, and the backhaul transmission of the inference results.
Accordingly, the latency for data offloading can be given by
\begin{equation}
    L_{\mathsf{VE}}^{\mathsf{off}} (t) = \frac{ D_{\mathsf{VE}} (t)}{r_{\mathsf{VE}}^{\mathsf{off}} (t)},
\end{equation}
where $D_{\mathsf{VE}} (t)$ is the offloading data volume of the MU $U_{\mathsf{VE}}$. Then, the latency for ES processing for the offloading data can be expressed as
\begin{equation}
    L_{\mathsf{VE}}^{\mathsf{ES}} (t) = \frac{\xi \chi D_{\mathsf{VE}}^{\mathsf{ES}} (t) + \zeta}
    {\omega_{\mathsf{VE}}(t)f_{\mathsf{ES}}},
\end{equation}
where $D_{\mathsf{VE}}^{\mathsf{ES}} (t) = \psi D_{\mathsf{VE}}^{\mathsf{}} (t)$ denotes the expected processed data at the ES with enhancement coefficient $\psi$, and $\omega_{\mathsf{VE}} (t)$ denotes the ratio of the data processing resources allocated to the MU $U_{\mathsf{VE}}$ at time slot $t$.
Moreover, the latency for the backhaul transmission of the inference results is thus given by
\begin{equation}
    L_{\mathsf{VE}}^{\mathsf{back}} (t) = \frac{D_{\mathsf{VE}}^{\mathsf{ES}} (t)}{r_{\mathsf{VE}}^{\mathsf{back}} (t)}.
\end{equation}

As a result, the overall latency of the MU $U_{\mathsf{VE}}$ is given by
\begin{equation}
    L_{\mathsf{VE}}^{\mathsf{}} (t) = L_{\mathsf{VE}}^{\mathsf{off}} (t) + L_{\mathsf{VE}}^{\mathsf{ES}} (t) + L_{\mathsf{VE}}^{\mathsf{back}} (t).
\end{equation}

\subsection{Problem Formulation}
In this letter, we aim to minimize the average of the overall latency of all MUs to improve the quality of heterogeneous services. Hence, the problem can be formulated as
\begin{subequations}\label{p1}
\begin{flalign}
  (\textbf{P1})\ \ &\min_{\boldsymbol{\alpha}, \boldsymbol\beta ,\boldsymbol\omega , \boldsymbol\lambda  }  \frac{1}{T}  \sum_{t\in \mathcal{T}} L_{\mathsf{comp}} (t)+L_{\mathsf{AIGC}} (t)+L_{\mathsf{VE}} (t)  \nonumber\\
 {\rm{s.t.}}  \quad 
&  \alpha_{\mathsf{comp}}^{\mathsf{off}}(t), \alpha_{\mathsf{VE}}^{\mathsf{off}}(t)  \in \left [ 0,1 \right ],  \quad \forall t\in \mathcal{T},\label{p1b}\\
& \alpha_{\mathsf{comp}}^{\mathsf{off}}(t)+ \alpha_{\mathsf{VE}}^{\mathsf{off}}(t)=1,  \quad \forall t\in \mathcal{T},\label{p1b}\\
& \alpha_{\mathsf{AIGC}}^{\mathsf{back}}(t), \alpha_{\mathsf{VE}}^{\mathsf{back}}(t)  \in \left [ 0,1 \right ],  \quad \forall t\in \mathcal{T},\label{p1d}\\
& \alpha_{\mathsf{AIGC}}^{\mathsf{back}}(t)+ \alpha_{\mathsf{VE}}^{\mathsf{back}}(t)=1,  \quad \forall t\in \mathcal{T},\label{p1e}\\
&  \beta \left ( t \right ), \lambda (t) \in \left [ 0,1 \right ], \quad \forall t\in \mathcal{T}, \label{p1f}\\
&  \omega_{\mathsf{comp}} (t), \omega_{\mathsf{AIGC}} (t), \omega_{\mathsf{VE}} (t) \in \left [ 0,1 \right ], \quad \forall t\in \mathcal{T}, \label{p1g}\\
& \omega_{\mathsf{comp}} (t) + \omega_{\mathsf{AIGC}} (t)+ \omega_{\mathsf{VE}} (t)=1, \quad \forall t\in \mathcal{T}, \label{p1h}
\end{flalign}
\end{subequations}
constraints \eqref{p1b}-\eqref{p1e} are the bandwidth allocations, constraint \eqref{p1f} denotes the power allocation for the ES backhaul transmission, and the offloading ratio of the MU $U_{\mathsf{comp}}$, respectively, and constraints \eqref{p1g}-\eqref{p1h} represent the computation resources allocation of the ES.

{Since the objective function of problem \textbf{P1} is non-convex, problem \textbf{P1} is a non-convex optimization problem. In addition, the optimization variables in \textbf{P1} are highly coupled. Hence, problem \textbf{P1} 
is hard to
solve by traditional optimization algorithms.
Fortunately,}
DRL can effectively address complex resource allocation challenges by dynamically adapting decisions to changing environments that include varying channels and diverse task requests \cite{8714026}. Therefore,
we propose a DRL-based latency-aware resource allocation (LARA) algorithm for solving problem \textbf{P1}.

\section{DRL-based Resource Allocation Algorithm}

It is observed that the problem \textbf{P1} can be reformulated as a Markov decision process (MDP), described by a tuple $\{s(t), a(t), r(t), \forall t\}$, where $s(t)$ denotes the system state, $a(t)$ represents the decision action, and $r(t)$ denotes the reward value for taking the action $a(t)$ at the state $s(t)$.

Specifically,
let $\mathcal{S}$ denote the state space, which
encompasses the channel conditions and task data sizes of all MUs. Hence, the state $s(t)$ at time slot $t$ is given by
\begin{equation}
    s(t) = [\boldsymbol{h}(t), D_{\mathsf{comp}}(t), D_{\mathsf{VE}}(t)],\ \forall t,
\end{equation}
where $\boldsymbol{h}(t) = \{h_{\mathsf{comp}}^{\mathsf{off}}(t), h_{\mathsf{AIGC}}^{\mathsf{back}}(t), h_{VE}^{\mathsf{off}}(t),h_{VE}^{\mathsf{back}}(t) \}$ includes the channel state information in the data offloading stage and the result backhaul transmission stage. Moreover, the
action space $\mathcal{A}$ consists of all decision variables in problem \textbf{P1}, and the action $a(t)$ at time slot $t$ can be represented as
\begin{equation}
    a(t) = [\boldsymbol{\alpha}(t) ,{\beta} \left ( t \right ), \lambda (t),\boldsymbol{\omega}(t)],\ \forall t,
\end{equation}
where ${\boldsymbol{\alpha}}(t)=[\alpha_{\mathsf{comp}}^{\mathsf{off}}(t), \alpha_{\mathsf{VE}}^{\mathsf{off}}(t), \alpha_{\mathsf{AIGC}}^{\mathsf{back}}(t), \alpha_{\mathsf{VE}}^{\mathsf{back}}(t)]$ denotes the bandwidth allocations, and $\boldsymbol{\omega}(t)=[\omega_{\mathsf{comp}} (t), \omega_{\mathsf{AIGC}} (t), \omega_{\mathsf{VE}} (t)]$ represents the computation resource allocations.
Since our goal is to minimize the overall latency of all MUs, we define the reward function as
\begin{equation}
    r(t) = -\frac{1}{T} \sum_{t\in \mathcal{T}} L_{\mathsf{comp}} (t)+L_{\mathsf{AIGC}} (t)+L_{\mathsf{VE}} (t),\ \forall t,
\end{equation}
which is the opposite of the objective function of \textbf{P1}.

Based on the above definition, we then use the deep deterministic policy gradient (DDPG) method to solve problem \textbf{P1} according to the MDP reformulation \cite{9214878}.
DDPG is a model-free off-policy actor-critic method designed for environments with continuous action spaces.
It
can approximate both the policy function (i.e., the actor function for giving actions) and the value function (i.e., the critic function that evaluates the action given by the actor function) to enable efficient and stable learning.
DDPG combines the deterministic policy gradient approach with the benefits of experience replay and target functions, providing robust performance in complex, high-dimensional tasks.
Specifically, the actor function is updated by minimizing the loss function
\begin{equation}
L_{\text{actor}} = -\mathbb{E}_{s \sim \rho^{\pi}} [Q(s, \mu(s|\theta^{\mu})|\theta^Q)],
\label{actorupdate}
\end{equation}
where 
$\theta^{\mu}$ and $\theta^{Q}$ are the parameters of the actor and critic functions, respectively, \(\mu(s|\theta^{\mu})\) is the actor's action under the policy $\rho^{\pi}$,
\(Q(s,a\vert \theta^{Q})\) is the evaluation of the critic function under state $s$, action $\mu(s|\theta^{\mu})$, and parameter $\theta^Q$. In addition, the critic network is updated by minimizing the loss function
\begin{equation}
L_{\text{critic}} = \mathbb{E}_{\{s,a,r,s'\} \sim {R}} [(y_i - Q(s_i,a_i|\theta^Q))^2],
\label{criticupdate}
\end{equation}
where $\{s,a,r,s'\}$ denotes the tuple sampled from the experience replay buffer \({R}\), $y_i$ represents the target evaluation value, which is given by
\begin{equation}
y_i = r + \gamma Q'(s', \mu'(s'|\theta^{\mu'})|\theta^{Q'}),
\end{equation}
where \(Q'\) and \(\mu'\) are the target critic and target actor functions, and \(\gamma\) is the discount factor. The utilization of experience replay and target functions helps stabilize training by breaking the correlation between consecutive updates and reducing the variance of the update target, respectively.
The proposed DDPG-based LARA algorithm is shown in Algorithm \ref{ddpg}.

\begin{algorithm} \scriptsize
\caption{LARA Algorithm for the MEGC System}
\label{ddpg}
\begin{algorithmic}[1]
\STATE Initialize actor network $\mu(s|\theta^\mu)$ and critic network $Q(s,a|\theta^Q)$; 
\STATE Initialize target networks $\theta^{\mu'} \leftarrow \theta^\mu$, $\theta^{Q'} \leftarrow \theta^Q$;
\STATE Initialize replay buffer $R$ and the maximum number of training episode $M$;
\FOR{episode = 1 to $M$}
    \STATE Initialize a random process $\mathcal{N}$ for action exploration;
    \STATE Receive initial observation state $s(1)$;
    \FOR{$t = 1$ to $T$}
        \STATE Select action $a(t) = \mu(s(t)|\theta^\mu) + \mathcal{N}_t$ according to the current policy and exploration noise;
        \STATE Execute action $a(t)$ and calculate reward $r(t)$ and new state $s(t+1)$;
        \STATE Store transition $(s(t), a(t), r(t), s(t+1))$ in replay buffer $R$;
        \STATE Sample a random minibatch of $N$ transitions $(s_n, a_n, r_n, s_{n+1})$ from $R$;
        \STATE Update critic function by minimizing the loss function \eqref{criticupdate}; 
        \STATE Update the actor function
        by minimizing the loss function \eqref{actorupdate};
        \STATE Update the target networks by soft update;
    \ENDFOR
\ENDFOR
\end{algorithmic}
\end{algorithm}

\textit{Complexity Analysis:}
In the LARA algorithm, the time complexity of each training step mainly includes the forward and backward propagation complexities of the actor and critic functions, as well as the complexity of updating the target functions. Therefore, the total time complexity of Algorithm 1 can be expressed as $\mathcal{O}(L_A \cdot n_A^2 + L_C \cdot n_C^2 + N_N)$, where $L_A$ and $L_C$ are the numbers of layers in the actor and critic functions, $n_A$ and $n_C$ are the number of neurons per layer, and $N_N$ is the number of network parameters. The testing time complexity is primarily determined by the forward propagation of the actor network, which is $\mathcal{O}(L_A \cdot n_A^2)$. Moreover, the total space complexity includes the storage requirements for the parameters and the experience replay buffer, which are $\mathcal{O}(2 \cdot (L_A \cdot n_A^2 + L_C \cdot n_C^2))$ and $\mathcal{O}(M_R \cdot (d_S + d_A + 1))$, respectively, where $M_R$ is the size of the experience replay buffer, $d_S$ and $d_A$ are the dimensions of the state space and action space.

\section{Numerical Results}
In this section, we test the performance of the proposed LARA algorithm by comparing it with four benchmarks: 1) fixed resource allocation (FRA) algorithm, where the optimization variables in \textbf{P1} are fixed; 2) random resource allocation (RRA) algorithm, where the optimization variables in \textbf{P1} are randomly selected in the feasible region;
{
3) discrete action based resource allocation (DARA) algorithm, where the optimization variables in \textbf{P1} are chosen from discrete actions, and the deep $\mathcal{Q}$-learning network is adopted;
4) exhaustive searching scheme (ESS), it discretizes the actions into multiple finely spaced discrete values and uses the exhaustive search algorithm to explore all possible solutions of these actions. By selecting the action that results in the minimum average time latency, the ESS effectively approaches the problem \textbf{P1}'s lower bound solution. Due to the adopted exhaustive search algorithm for selecting actions, the ESS has high complexity\cite{9417469,9136602}.}
The distances between the ES and the MU $U_{\mathsf{comp}}$, the MU $U_{\mathsf{AIGC}}$, and the MU $U_{\mathsf{VE}}$, are $100,120,80$ meters, respectively.
The transmit power of the MUs are $p_{\mathsf{comp}} = p_{\mathsf{VE}} = 15$W, and the system bandwidth is $B_{\mathsf{off}} = B_{\mathsf{back}} = 400$MHz.
The transmit power of the ES is $P=15$W.
The offloading task data volume of the MUs $U_{\mathsf{comp}}$ and $U_{\mathsf{VE}}$ follows a uniform distribution of $[1,8]$ Mbits.
The noise power density is $N_0 = -100$dBm/Hz.
The GPT-4 is employed as the AIGC model \cite{10578004},
and the generation-related parameters $\xi$ and $\zeta$ are $9.97 \times 10^{-14}$ and $5.73$, respectively.
The generated data volume $D_{\mathsf{AIGC}}^{\mathsf{ES}}$ and $D_{\mathsf{VE}}^{\mathsf{ES}}$ are taken from $[2,16]$ Mbits according to specific AIGC and VE requests, respectively.

\begin{figure}[!h]
	\begin{minipage}{0.475\linewidth}
		\centering
		\includegraphics[width=1\linewidth]{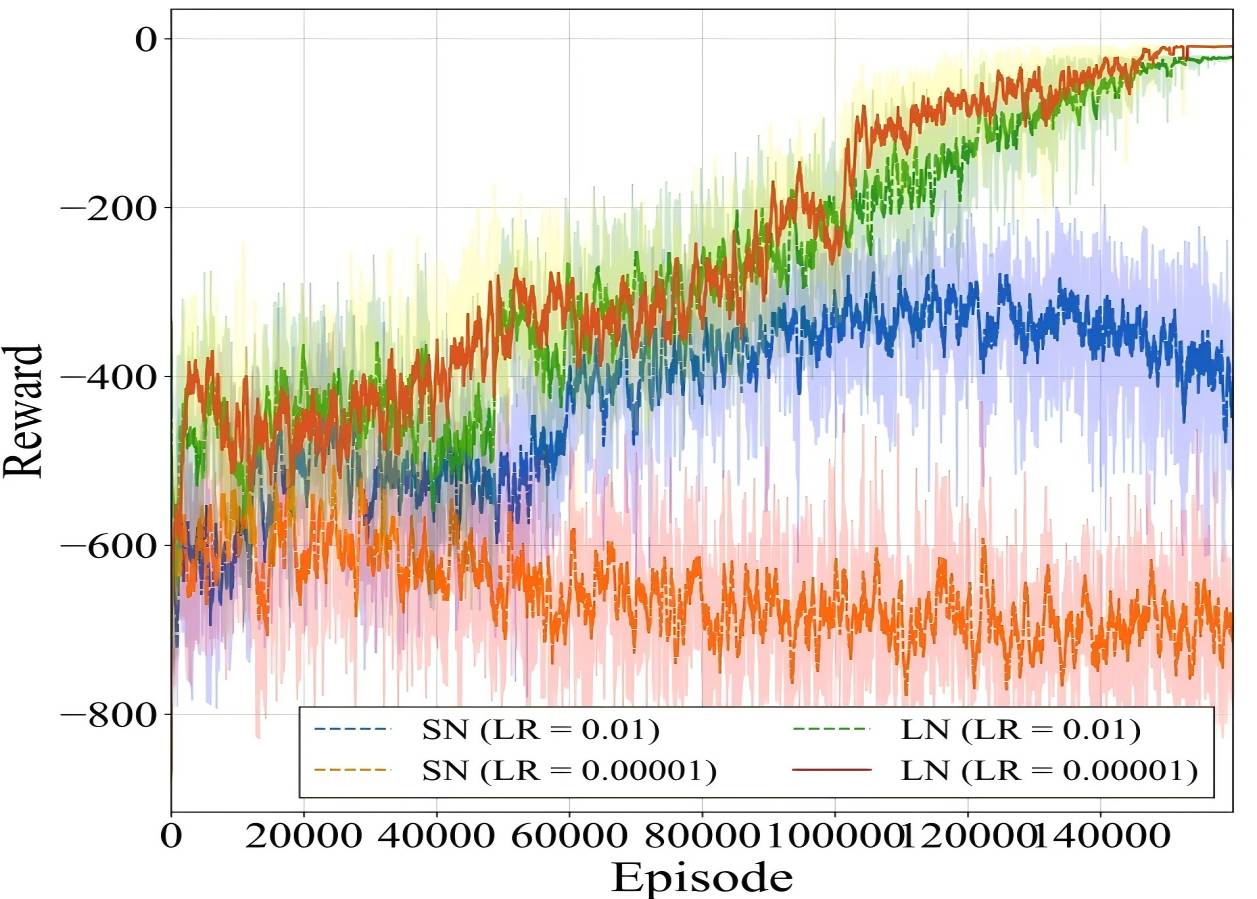}
		\captionsetup{font=small}
            \caption{\centering{The reward value versus the training episodes under different learning rates.
            }}
		\label{reward_curve}
	\end{minipage}
	\begin{minipage}{0.465\linewidth}
		\centering
		\includegraphics[width=1\linewidth]{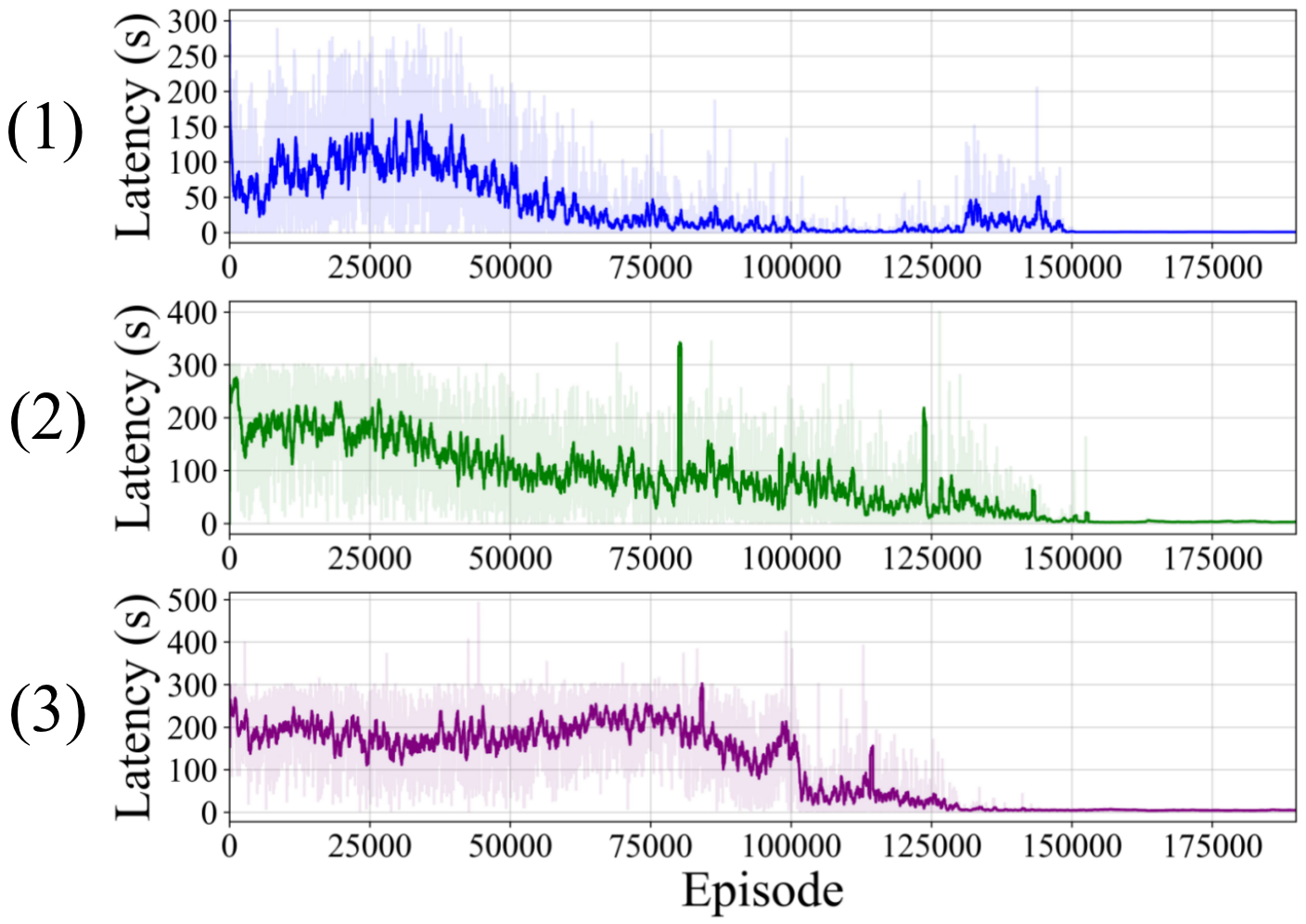}
            \captionsetup{font=small}
		\caption{\centering{Latency versus the training episodes: (1) the MU $U_{\mathsf{comp}}$, (2) the MU $U_{\mathsf{AIGC}}$, (3) the MU $U_{\mathsf{VE}}$.
            }}
		\label{figuser}
	\end{minipage}
\end{figure}


{Fig. \ref{reward_curve} illustrates the training process of the proposed LARA algorithm under different learning rates (LRs) for both large network (LN) and small network (SN) structures. 
The LN-based LARA comprises $5$ hidden layers, and the maximum number of neurons in each layer is $4,096$. In contrast, the SN-based LARA consists of $4$ hidden layers, and each layer has a maximum number of $600$ neurons. It is evident that the LN-based LARA achieves better convergence rewards and maintains stability during training. The red curve indicates that when the LARA has large network structure, a smaller LR can more closely approximate the optimal policy. Conversely, the orange curve demonstrates that when the network has inadequate hidden layers and neurons, a smaller LR may result in slow policy updates, leading to a suboptimal strategy.}
Fig. \ref{figuser} shows the latency performance versus the training episodes for the three MUs.
Despite each MU having distinct service requirements and varying communication and computation resources,
the latency of each MU eventually converges.
This observation validates the effectiveness and adaptability of the proposed LARA algorithm in tackling complex and heterogeneous resource allocation problems.



\begin{figure}[!h]
	\begin{minipage}{0.495\linewidth}
		\centering
		\includegraphics[width=1\linewidth]{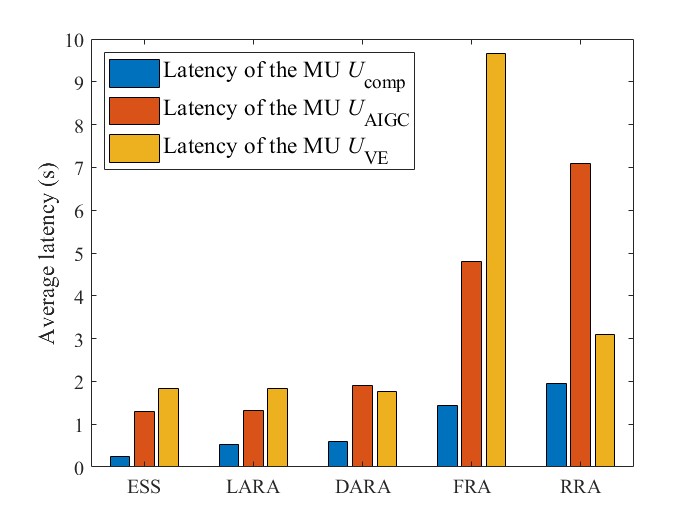}
		\captionsetup{font=small}
            \caption{\centering{The average latency vs. different algorithms.
            }}
		\label{simulation}
	\end{minipage}
	\begin{minipage}{0.495\linewidth}
		\centering
		\includegraphics[width=1\linewidth]{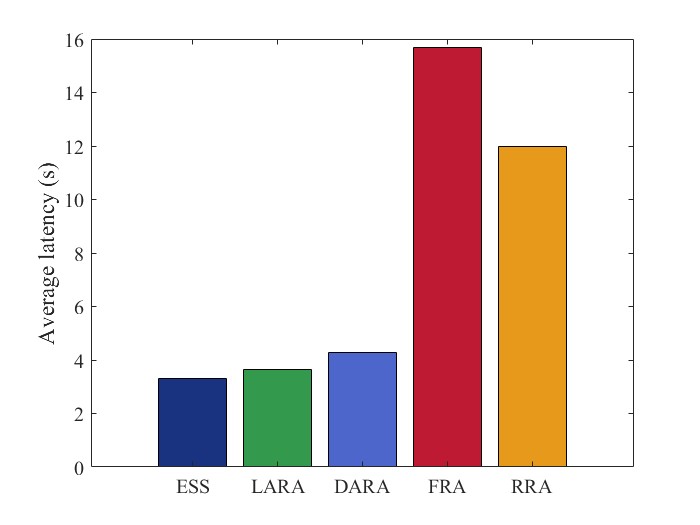}
            \captionsetup{font=small}
		\caption{\centering{The total latency vs. different algorithms.
            }}
		\label{user_sum}
	\end{minipage}
\end{figure}

Fig. \ref{simulation} and Fig. \ref{user_sum} show the latency of each MU and the total latency for all MUs under the proposed LARA algorithm and two benchmark algorithms.
The proposed LARA algorithm can achieve the smallest latency among all the algorithms, {and has near performance to the ESS algorithm.}
The reason is that the proposed
LARA algorithm is capable of adjusting the action according to the time-varying state information {in a continuous action selection manner}.
Furthermore, each MU's latency performance in the proposed LARA algorithm achieves the lowest latency compared to the corresponding MU in the four benchmarks, {and also close to the ESS.}
The fairness of the proposed LARA algorithm to provide services to all MUs at the same time is thus verified.
In addition, we can observe that the latency of the MU $U_{\mathsf{comp}}$ is the lowest among the three MUs.
The reason is that the MU $U_{\mathsf{comp}}$ does not request AIGC or VE services at the ES,
while the backhaul transmission latency for the computation results can be ignored.
In contrast, the MUs $U_{\mathsf{AIGC}}$ and $U_{\mathsf{VE}}$ require the computation resources of the ES for the AIGC/VE services, and also require the backhaul transmissions for the results with an assignable amount of data volume.

\section{Conclusion}
This letter investigate
the latency minimization problem in an MEGC system to provide computation, AIGC, and VE services for MUs.
The bandwidth allocation, the backhaul transmit power, the computation resources, and the task offloading ratio, are jointly optimized. A novel DRL-based LARA algorithm is designed to solve the optimization problem.
Finally,
simulation results
demonstrate that the proposed LARA algorithm outperforms the baseline algorithms in terms of the latency performance.
{Besides, the selections of the LR and network structure are also analyzed.}

\bibliographystyle{IEEEtran}
\bibliography{ref}

\begin{thebibliography}{10}
\providecommand{\url}[1]{#1}
\csname url@samestyle\endcsname
\providecommand{\newblock}{\relax}
\providecommand{\bibinfo}[2]{#2}
\providecommand{\BIBentrySTDinterwordspacing}{\spaceskip=0pt\relax}
\providecommand{\BIBentryALTinterwordstretchfactor}{4}
\providecommand{\BIBentryALTinterwordspacing}{\spaceskip=\fontdimen2\font plus
\BIBentryALTinterwordstretchfactor\fontdimen3\font minus \fontdimen4\font\relax}
\providecommand{\BIBforeignlanguage}[2]{{%
\expandafter\ifx\csname l@#1\endcsname\relax
\typeout{** WARNING: IEEEtran.bst: No hyphenation pattern has been}%
\typeout{** loaded for the language `#1'. Using the pattern for}%
\typeout{** the default language instead.}%
\else
\language=\csname l@#1\endcsname
\fi
#2}}
\providecommand{\BIBdecl}{\relax}
\BIBdecl

\bibitem{10398474}
M.~Xu, H.~Du, D.~Niyato, J.~Kang, Z.~Xiong, S.~Mao, Z.~Han, A.~Jamalipour, D.~I. Kim, X.~Shen, V.~C.~M. Leung, and H.~V. Poor, ``Unleashing the power of edge-cloud generative {AI} in mobile networks: A survey of {AIGC} services,'' \emph{IEEE Commun. Surveys Tuts.}, vol.~26, no.~2, pp. 1127--1170, 2nd Quart. 2024.

\bibitem{10578004}
R.~Zhong, X.~Mu, Y.~Zhang, M.~Jaber, and Y.~Liu, ``Mobile edge generation: A new era to {6G},'' \emph{IEEE Network}, vol.~38, no.~5, pp. 47--55, Sep. 2024.

\bibitem{xu2024}
X.~Xu, Y.~Liu, X.~Mu, H.~Xing, and A.~Nallanathan, ``Accelerating mobile edge generation ({MEG}) by constrained learning,'' \emph{arXiv:2407.07245}, 2024.

\bibitem{8016573}
Y.~Mao, C.~You, J.~Zhang, K.~Huang, and K.~B. Letaief, ``A survey on mobile edge computing: The communication perspective,'' \emph{IEEE Commun. Surveys Tuts.}, vol.~19, no.~4, pp. 2322--2358, 4th Quart. 2017.

\bibitem{8664595}
J.~Ren, G.~Yu, Y.~He, and G.~Y. Li, ``Collaborative cloud and edge computing for latency minimization,'' \emph{IEEE Trans. Veh. Technol.}, vol.~68, no.~5, pp. 5031--5044, May 2019.

\bibitem{8387798}
J.~Ren, G.~Yu, Y.~Cai, and Y.~He, ``Latency optimization for resource allocation in mobile-edge computation offloading,'' \emph{IEEE Trans. Wireless Commun.}, vol.~17, no.~8, pp. 5506--5519, Aug. 2018.

\bibitem{10472660}
J.~Wang, H.~Du, D.~Niyato, J.~Kang, Z.~Xiong, D.~Rajan, S.~Mao, and X.~Shen, ``A unified framework for guiding generative {AI} with wireless perception in resource constrained mobile edge networks,'' \emph{IEEE Trans. Mobile Comput.}, vol.~23, no.~11, pp. 10\,344--10\,360, Nov. 2024.

\bibitem{xu20242}
X.~Xu, R.~Zhong, X.~Mu, Y.~Liu, and K.~Huang, ``Mobile edge generation-enabled digital twin: Architecture design and research opportunities,'' \emph{arXiv:2407.02804}, 2024.

\bibitem{10319405}
Z.~He, Y.~Sun, B.~Wang, S.~Li, and B.~Zhang, ``{CPU–GPU} heterogeneous computation offloading and resource allocation scheme for industrial internet of things,'' \emph{IEEE Internet Things J.}, vol.~11, no.~6, pp. 11\,152--11\,164, Mar. 2024.

\bibitem{10535750}
H.~Ding, Z.~Zhao, H.~Zhang, W.~Liu, and D.~Yuan, ``{DRL} based computation efficiency maximization in {MEC}-enabled heterogeneous networks,'' \emph{IEEE Trans. Veh. Technol.}, to appear in 2024.

\bibitem{10606316}
W.~Liu, H.~Wang, X.~Zhang, H.~Xing, J.~Ren, Y.~Shen, and S.~Cui, ``Joint trajectory design and resource allocation in {UAV}-enabled heterogeneous {MEC} systems,'' \emph{IEEE Internet Things J.}, vol.~11, no.~19, pp. 30\,817--30\,832, Oct. 2024.

\bibitem{9557844}
C.-L. Chen, C.~G. Brinton, and V.~Aggarwal, ``Latency minimization for mobile edge computing networks,'' \emph{IEEE Trans. Mobile Comput.}, vol.~22, no.~4, pp. 2233--2247, Apr. 2023.

\bibitem{8714026}
N.~C. Luong, D.~T. Hoang, S.~Gong, D.~Niyato, P.~Wang, Y.-C. Liang, and D.~I. Kim, ``Applications of deep reinforcement learning in communications and networking: A survey,'' \emph{IEEE Commun. Surveys Tuts.}, vol.~21, no.~4, pp. 3133--3174, 4th Quart. 2019.

\bibitem{9214878}
Y.~Chen, Z.~Liu, Y.~Zhang, Y.~Wu, X.~Chen, and L.~Zhao, ``Deep reinforcement learning-based dynamic resource management for mobile edge computing in industrial internet of things,'' \emph{IEEE Trans. Ind. Informat.}, vol.~17, no.~7, pp. 4925--4934, Jul. 2021.

\bibitem{9417469}
X.~Zhang, Y.~Shen, B.~Yang, W.~Zang, and S.~Wang, ``Drl based data offloading for intelligent reflecting surface aided mobile edge computing,'' in \emph{Proc. IEEE Wireless Commun. Networking Conf., WCNC}, Nanjing, China, Mar. 2021, pp. 1--7.

\bibitem{9136602}
S.~Gong, Y.~Xie, J.~Xu, D.~Niyato, and Y.-C. Liang, ``Deep reinforcement learning for backscatter-aided data offloading in mobile edge computing,'' \emph{IEEE Network}, vol.~34, no.~5, pp. 106--113, Sep./Oct. 2020.

\end{thebibliography}
\end{document}